\documentclass[sigconf, nonacm]{acmart}

\usepackage{graphicx}
\usepackage{hyperref}
\usepackage{booktabs}
\usepackage[table, dvipsnames]{xcolor}
\usepackage{tabularx}
\usepackage{multirow}
\usepackage{ragged2e}
\usepackage{subcaption}
\usepackage{balance}
\usepackage{enumitem}

\usepackage{tikz}
\usepackage{tcolorbox}
\tcbuselibrary{skins, breakable}
\usepackage{soul}
 
\definecolor{colA}{HTML}{2166AC}   
\definecolor{colB}{HTML}{C1272D}   
\definecolor{colC}{HTML}{1B7837}   
\definecolor{colD}{HTML}{B8860B}   
\definecolor{colE}{HTML}{7B3294}   
 
\definecolor{bgA}{HTML}{D1E5F0}    
\definecolor{bgB}{HTML}{F8C8CC}    
\definecolor{bgC}{HTML}{D9F0D3}    
\definecolor{bgD}{HTML}{FFF2B2}    
\definecolor{bgE}{HTML}{E7D4EF}    
 
\newcommand{\hlA}[1]{{\sethlcolor{bgA}\hl{#1}}}
\newcommand{\hlB}[1]{{\sethlcolor{bgB}\hl{#1}}}
\newcommand{\hlC}[1]{{\sethlcolor{bgC}\hl{#1}}}
\newcommand{\hlD}[1]{{\sethlcolor{bgD}\hl{#1}}}
\newcommand{\hlE}[1]{{\sethlcolor{bgE}\hl{#1}}}
 
\soulregister{\textbf}{1}
\soulregister{\texttt}{1}
\soulregister{\emph}{1}
 
\newcommand{\swatch}[1]{%
  \tikz[baseline=-0.6ex]{%
    \fill[#1, rounded corners=1pt] (0,0) rectangle (0.7em,0.7em);}}
 
\AtBeginDocument{%
  }

\copyrightyear{2025}
\acmYear{2025}
\setcopyright{rightsretained}
\acmConference[TODO '25]{TODO 25': TODOTODO}{MMMM DD--DD, 2025}{City, Country}
\acmBooktitle{TODO}
\acmPrice{}
\acmDOI{TODO}
\acmISBN{TODO}

\begin{document}


\title{Exploring Emerging Norms of AI Attribution and Disclosure in Programming Education}

\author{Runlong Ye}
\orcid{0000-0003-1064-2333}
\email{harryye@cs.toronto.edu}
\affiliation{
  \institution{Computer Science, \\University of Toronto}
  \city{Toronto}
  \state{Ontario}
  \country{Canada}
}

\author{Oliver Huang}
\orcid{0009-0007-1585-1229}
\email{oliver@cs.toronto.edu}
\affiliation{
  \institution{Computer Science, \\University of Toronto}
  \city{Toronto}
  \state{Ontario}
  \country{Canada}
}

\author{Jessica He}
\orcid{0000-0003-2368-0099}
\email{jessicahe@ibm.com}
\affiliation{
 \institution{IBM Research}
 \city{Cambridge}
 \state{Massachusetts}
 \country{USA}
}

\author{Michael Liut}
\orcid{0000-0003-2965-5302}
\email{michael.liut@utoronto.ca}
\affiliation{
  \institution{Mathematical and Computational Sciences, \\University of Toronto Mississauga}
  \city{Mississauga}
  \state{Ontario}
  \country{Canada}
}


\begin{abstract}
Generative AI blurs the lines of authorship in computing education, creating uncertainty around how students should attribute AI assistance. To examine these emerging norms, we conducted a factorial vignette study with 94 computer science students across 102 unique scenarios, systematically manipulating assessment type, AI autonomy, student activity, prior knowledge, and human refinement effort. This paper details how these factors influence students' perceptions of ownership and disclosure preferences. Our findings indicate that attribution judgments are primarily driven by different levels of AI assistance and human refinement. We also found that students' perception of authorship significantly predicts their policy expectations. We conclude by proposing a shift from statement-style policies to process-oriented attribution, transforming disclosure into a pedagogical mechanism for fostering critical engagement with AI-generated content.
\end{abstract}




\maketitle

\vspace{-1.5em}
\section{Introduction}


Generative AI has been shown to offer immediate performance gains in various academic settings, including programming education, allowing novices to complete more exercises and score higher \cite{kazemitabaar2023studying, akccapinar2024ai}. Yet, these gains often come at the expense of durable skill building; without pedagogical guardrails, reliance on AI solutions can reduce future performance and self-efficacy \cite{bastani2025generative, lee2025impact, margulieux2024self}. While attribution is a standard mechanism for academic integrity \cite{park2017other}, it also offers a pathway to responsible use by requiring students to critically evaluate and explicitly delineate their contribution \cite{howard1992plagiarism}. However, norms for AI disclosure remain contested. Universities and publishers struggle with inconsistent definitions of ``originality'' and unreliable detection tools \cite{luo2024critical, resnik2025disclosing, weber2023testing}, shifting the burden from technical compliance to individual student judgment. To design effective guardrails, we must first understand how students perceive ownership in these workflows.

Critically, existing research on attribution focuses on prose and creative writing \cite{he2025contributions, he2025exploring}, overlooking the unique functional nature of computing. In programming, contributions are modular and executable; ``ownership'' often associates with debugging, integration, and testing rather than just code generation. It remains unclear how standard authorship norms translate to these technical tasks. 

To address this gap and explore the potential of attribution as a pedagogical guardrail, we are interested in how students decide when, how, and how much to attribute AI assistance across various programming assessments. Specifically, we aim to answer the following research questions (RQs): 
\vspace{-.4em}
\begin{enumerate}
    \item[\textbf{RQ1}:] How do different dimensions of student-AI workflows influence students' attribution perceptions and disclosure preferences?
    \begin{enumerate}
            \item \textit{(Perception):} How do these dimensions shape internal judgments of authorship, initiative, and learning?
            \item \textit{(Preference):} How do these dimensions drive students' preferred disclosure methods?
        \end{enumerate}
    \item[\textbf{RQ2}:] How do students' internal perceptions of authorship and pedagogical value predict preferred disclosure methods?
\end{enumerate}
\vspace{-.4em}

To answer these questions, we conducted a factorial vignette study with 94 computer science students across 102 scenarios, systematically manipulating five unique dimensions, and asked participants to rate credit allocation and disclosure requirements after reading these scenarios. Our analysis revealed that among these factors, AI assistance level and human effort emerged as the primary drivers of attribution judgments. These findings propose a shift toward \textit{process-oriented attribution}, requiring students to demonstrate thoughtful engagement with AI, aiming to transform AI disclosure into a source of critical engagement and productive friction.

\section{Related Work}
\subsection{Attribution as a Pedagogical Guardrail}
Self-regulated learning (SRL) is foundational to computing and programming education \cite{falkner2014identifying}, yet Generative AI introduces significant metacognitive risks. While AI tools can scaffold performance, poorly supported engagement often leads to over-trust and an ``illusion of competence'' where students produce correct code without understanding~\cite{prather2024widening, akccapinar2024ai, margulieux2024self}. To mitigate this, researchers have developed various pedagogical guardrails, from cognitive forcing functions that compel reflection~\cite{buccinca2021trust, kazemitabaar2025exploring, kumar2024guiding} to constrained environments~\cite{paludo2024fostering} that require critical evaluation of AI outputs.

We propose that \textit{attribution}, the act of explicitly distinguishing AI's contribution from one's own, can serve as a similar metacognitive guardrail. Unlike technical restrictions or watermarking, attribution requires the student to actively delineate their agency. However, while SRL frameworks in computing are well-established~\cite{loksa2016role, loksa2022metacognition}, the specific norms for how attribution should function as a learning mechanism remain unexplored. We investigate the underlying perceptions that would inform such a guardrail, specifically examining how varying levels of AI assistance and human refinement shape students' judgments of ownership and learning value.

\subsection{Attribution Practices and The Gap in Programming Education}
Current attribution norms are largely derived from creative and scientific writing, where guidelines focus on transparency and liability. Policies from model providers and academic journals mandate disclosure but generally reject AI authorship on the grounds of accountability~\cite{openai2022sharing, springernature_ai_nd}. Empirical studies in co-creative writing reveal that users assign less credit to AI than to humans for equivalent work, suggesting that attribution is deeply contextual rather than rule-based~\cite{he2025contributions, he2025exploring}. However, research in this area is largely rooted in prose, where ``authorship'' is defined by textual generation~\cite{draxler2024ai, fang2025shapes, hwang202580}.

This text-centric view fails to capture the unique nature of programming, where contributions are functional, modular, and executable. In programming, ``ownership'' often hinges on integration, debugging, and testing rather than just line-by-line generation. It remains unclear how students navigate these distinctions; for instance, does verifying AI-generated code constitute ownership? Our work bridges this gap by empirically examining how the mechanics of programming assistance (e.g., autonomy, refinement effort) drive attribution judgments, moving beyond the binary notions of originality found in current policy.
\section{Methods}
We conducted a vignette-based factorial survey with 94 computer science students. To capture the diversity of computing education, we systematically manipulated five factors, including \textbf{A) Assessment Type}, \textbf{B) AI Assistance Level}, \textbf{C) Activity Type}, \textbf{D) Prior Knowledge}, and \textbf{E) Human Post-AI Effort}, resulting in 102 unique scenario combinations. We selected these dimensions to isolate how the specific mechanics of the programming workflow, from the tool's autonomy to the student's refinement effort, shape perceptions of ownership across realistic educational contexts. We curated the list of factors through iterative refinement, balancing the need to comprehensively model diverse student-AI workflows with the practical need to keep the experimental design manageable. Factor definitions and factorial design details are provided in the Appendix \ref{appendix:factorial}.

\paragraph{Procedure and Materials.}
Recruitment was conducted via Prolific. Participants were screened to ensure they were currently enrolled in or recently graduated from a computing-related program. The final sample ($N=94$) spanned diverse academic levels, from first-year undergraduates to final-year/recent graduates.

To minimize cognitive burden, vignettes were presented as interactive HTML slide decks (piloted with $N=10$ for clarity) in addition to text blocks. Each vignette displayed: (1) the assessment context; (2) a concrete description of the AI assistance; (3) the student's integration process; and (4) a snapshot of the deliverable. 

Participants first completed a baseline questionnaire on their exposure to AI policies and attribution habits. They then evaluated 5 vignettes drawn via randomized sampling to ensure uniform coverage of the factor conditions. For each vignette, they provided judgments on credit allocation and disclosure requirements. The session concluded with open-ended reflection questions regarding their decisions. The study took approximately 30 minutes and compensated at the Prolific-recommended rate of \$12/hr.

\paragraph{Dependent Measures.}
We measured student perceptions across three distinct categories:
\begin{itemize}[leftmargin=*]
\item \textbf{Perceived Authorship and Agency:} Four 7-point scales assessing the locus of control on a spectrum from ``Entirely Human'' (1) to ``Entirely AI'' (7). Specific items measured \textit{Work Attribution} (execution effort), \textit{Intellectual Contribution} (new ideas), \textit{Initiative} (advancing the work), and \textit{Liability} (responsibility for errors).
\item \textbf{Pedagogical Validity:} Two 7-point scales evaluating the educational integrity of the scenario, including \textit{Grading Fairness} and \textit{Student Learning}.
\item \textbf{Disclosure Preferences:} Two items measuring external demands versus personal intent. \textit{Attribution Intensity} asked what policy \textit{should} be required (\textit{None}, \textit{Brief acknowledgment}, \textit{Citation}, or \textit{Co-authorship} \cite{luo2024critical, resnik2025disclosing}), while \textit{Personal Willingness} captured whether the student would \textit{personally} disclose AI use (Yes/No).
\end{itemize}

\paragraph{Data Analysis.}
To analyze internal perceptions (RQ1a), we used Type II Analysis of Variance (ANOVA) on the Likert metrics, calculating partial eta-squared ($\eta_p^2$) for effect sizes. To analyze disclosure preferences (RQ1b), we used Pearson’s Chi-Square ($\chi^2$) tests. Finally, to determine how internal beliefs drive policy demands (RQ2), we modeled the relationship between perception scores and disclosure requirements using Ordered Logistic Regression.

\section{Results}

\subsection{RQ1a: Perceptions of Authorship rely on AI Autonomy and Human Refinement}
\label{sec:result-rq1a}

\begin{figure*}[ht]
    \centering
    \includegraphics[width=.85\linewidth]{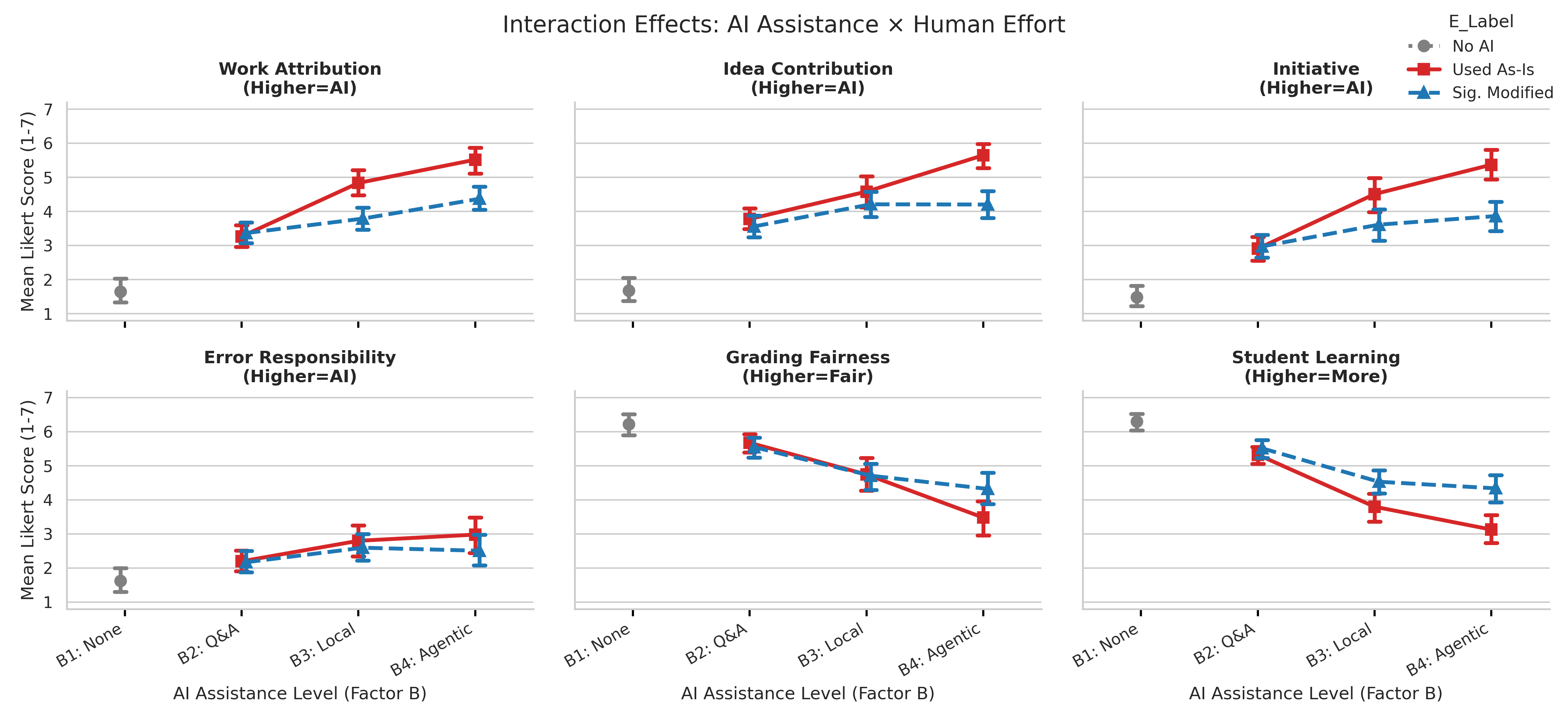}
    \caption{Interaction between AI assistance level (B1: None, B2: Q\&A, B3: Local generation, B4: Systemic/agentic) and human post-AI effort (E1: Used As-Is, E2: Significantly Modified) on six student-perception outcomes; points show mean Likert ratings (1–7) and error bars show standard error (SEM). No AI (B1) is shown as a baseline.}
    \label{fig:B-E-interaction}
\end{figure*}

To determine which factors drive students' internal assessments of work and liability, we analyzed the five experimental factors using factorial ANOVA. Attribution judgments were driven primarily by the mechanics of the human-AI interaction rather than the academic context. The \textbf{AI Assistance Level (Factor B)} emerged as the strongest and most consistent predictor across all dimensions ($p < .001$). It accounted for the largest proportion of variance in judgments of Work Attribution ($\eta_p^2 = 0.16$) and Initiative ($\eta_p^2 = 0.14$). As the AI progressed from simple Q\&A to agentic generation, students linearly discounted their own authorship and initiative. As illustrated in the effect size heatmap (Appendix Table \ref{tab:anova_main_effects}).

\textbf{Human Post-AI Effort (Factor E)} was the secondary driver, though its impact was notably smaller than that of AI assistance. Human effort had a statistically significant but modest effect on Work Attribution ($\eta_p^2 = 0.04$) and Student Learning ($\eta_p^2 = 0.05$). Other factors like \textit{Assessment Type (A)}, \textit{Activity Type (C)}, and \textit{Prior Knowledge (D)}, had near-zero effect sizes ($\eta^2 \approx 0.00$).

The interaction plots between Factor B (AI Assistance Level) and E (Post-AI Effort) in Figure \ref{fig:B-E-interaction} reveal that while modifying AI-generated code skewed student perception compared to the \textit{Used As-Is} condition, this restorative effect was insufficient to counteract the dominance of higher-level AI assistance. Where AI acted as an autonomous agent, students hesitated to claim full ownership of ideas and initiative, even when they invested notable effort in refinement.

Notably, other factors like \textit{Assessment Type (A)}, \textit{Activity Type (C)}, and \textit{Prior Knowledge (D)}, had near-zero effect sizes ($\eta^2 \approx 0.00$). Students applied the same internal logic regarding authorship and learning whether the task was a capstone project or a weekly lab, and whether they were planning concepts or implementing code.

\subsection{RQ1b: Higher AI Autonomy and Implementation Tasks Trigger Stricter Disclosure Expectation}
\label{sec:result-rq1b}

\begin{figure}[ht]
    \centering
    \includegraphics[width=.8\linewidth]{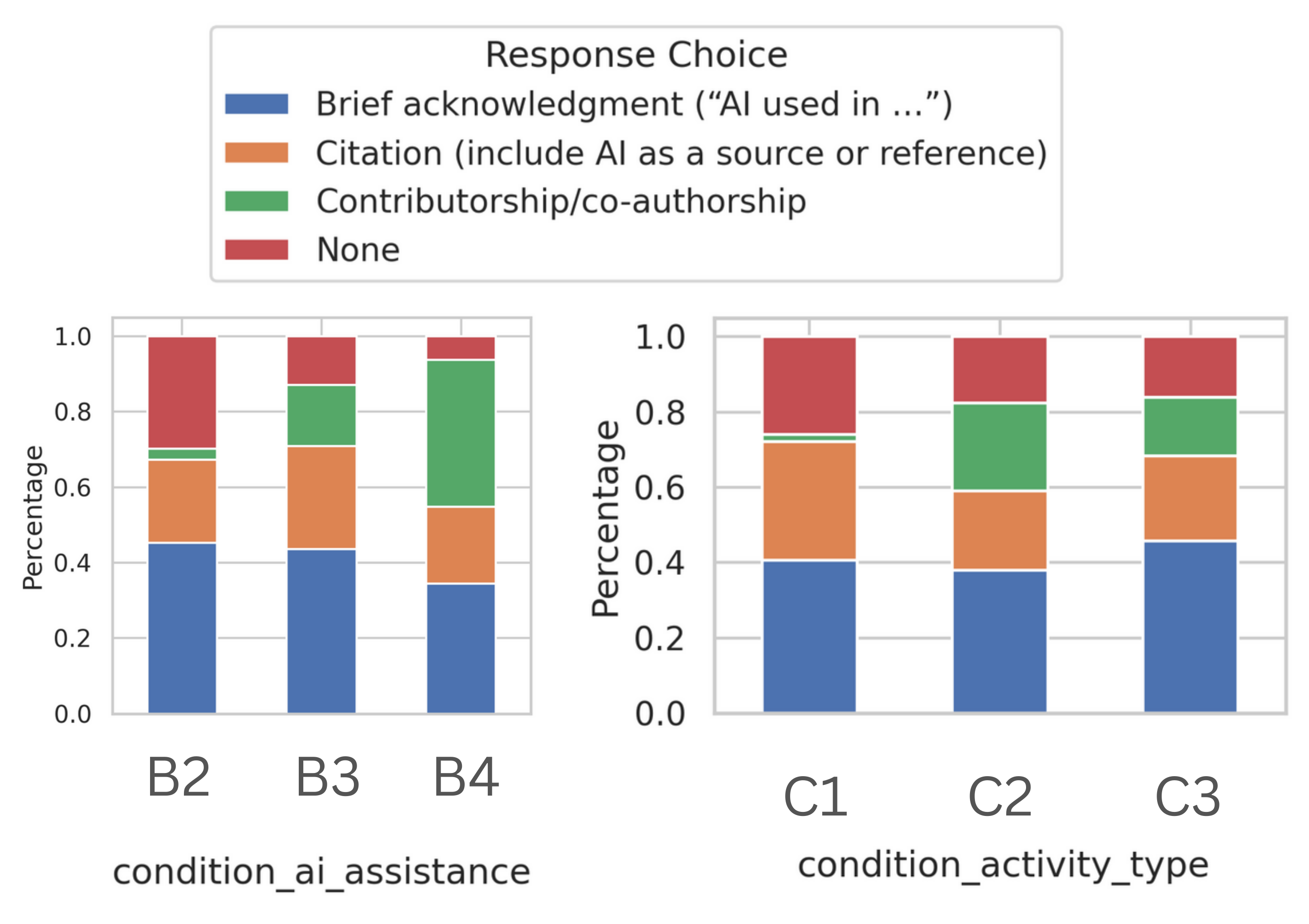}
    \caption{Distribution of disclosure preferences by AI Assistance Level (Left) and Activity Type (Right). AI assistance (Factor B) drives a shift towards ``Co-authorship,'' while Production tasks (Factor C) trigger stricter citation requirements than Planning. We omit No AI (B1) because disclosure is trivially `None' in that condition.}
    \label{fig:combined-disclosure-pref}
\end{figure}

Our analysis of categorical disclosure preferences reveals that students’ ownership judgments are largely driven by interaction mechanics (\textit{AI assistance (A)} and \textit{post-AI effort (E)}), while disclosure expectations also vary with \textit{activity type (C)}.

\textbf{AI Assistance Level (Factor B)} was the dominant predictor of disclosure behavior ($\chi^2 = 202.40, p < .001$). As illustrated in Figure \ref{fig:combined-disclosure-pref} (left), preferences shifted distinctively across levels: starting with Q\&A support (B2), students' consensus is to ``Brief Acknowledgment'' (blue). Notably, the \textit{Systemic/Agentic} condition (B4) triggered a surge in ``Co-authorship'' (green), indicating a threshold at which autonomous code generation transforms the perceived relationship from tool usage to human-AI collaboration and thus warrants a more predominant form of attribution.

\textbf{Activity Type (Factor C)} also significantly modulated disclosure preference ($\chi^2 = 21.24, p = .002$). As seen in Figure \ref{fig:combined-disclosure-pref} (right), tasks involving \textit{Production/Implementation} (C2) demanded stricter formal citation compared to \textit{Conceptual Planning} (C1). This suggests that the generation of executable artifacts triggers stronger transparency expectations than the generation of abstract ideas.

In contrast, \textit{Human Post-AI Effort (Factor E)} ($p = .303$), \textit{Assessment Type (A)} ($p = .194$), and \textit{Prior Knowledge (D)} ($p = .080$) did not significantly influence disclosure preferences. While our previous results showed that human effort restores internal perceptions of ownership, these findings indicate that it does not excuse students from the external obligation to disclose use of AI.

\subsection{RQ2: Authorship Concerns Drive Policy Requirements, But Do Not Predict Personal Intent to Disclose}
\label{sec:result-rq2}
To understand the cognitive basis of AI attribution, we modeled the relationship between students' internal perceptions of the scenario (e.g., ``Who did the work?'', ``Who is responsible for errors?'') and their attribution decisions. We performed two separate Ordered Logistic Regressions to distinguish between \textbf{Policy Requirements} (what policy \textit{should} be mandated) and \textbf{Personal Intent} (what students would \textit{actually} do).

\subsubsection{Policy Requirements are Driven by Authorship, Not Liability}
We first examined what drives students to demand mandatory disclosure policies (ranging from ``None'' to ``Co-authorship''). The regression analysis reveals that students view attribution mandates primarily as a mechanism for assigning intellectual credit rather than assigning technical liability. 

First, we observed a strong positive relationship between perceived AI contribution and disclosure stringency. \textbf{Work Attribution} ($\beta = 0.28, p = .009$) and \textbf{Idea Contribution} ($\beta = 0.25, p = .011$) were significant positive predictors. This confirms that as students perceive the labor and ideation to be more ``AI-generated'' (higher Likert scores), they demand significantly higher levels of formal acknowledgment. This aligns with standard authorship norms that credit must follow contribution.

Second, and perhaps more revealing, \textbf{Grading Fairness} emerged as the strongest negative predictor ($\beta = -0.38, p < .001$). This negative coefficient indicates an inverse relationship: when participants viewed a submission as ``fair to grade'' (implying the AI use was legitimate and within the bounds of the assignment), the demand for strict disclosure decreased. Conversely, when AI use was perceived as creating an \textit{unfair} grading situation, signaling a breach of academic integrity, participants demanded rigorous, formal disclosure. Full regression results can be found in Appendix Table \ref{tab:disclosure_req_result}.

\subsubsection{Personal Intention Diverges from Policy}
When we shifted the analysis to predict \textit{Personal Practice} (i.e., ``Would you personally disclose AI use?''), the explanatory power of the model collapsed. In contrast to policy requirements, none of the six internal perception variables significantly predicted students' personal practice to disclose ($p > .10$ for all factors, see Appendix Table \ref{tab:disclosure_pref_result}). While students consider institutional rules \textit{should} be calibrated based on \textit{Work}, \textit{Ideas}, and \textit{Fairness in Grading}, their own intended behavior does not mirror these dimensions. This suggests that personal disclosure decisions are likely driven by latent, unmeasured factors, such as risk aversion, compliance habits, or fear of academic penalties, rather than an objective evaluation of the specific pedagogical or intellectual context.
\section{Discussion}
\paragraph{Attribution Norms: Interaction Mechanics vs. Assessment Type}
Our findings suggest that students' judgments of ownership are primarily driven by the mechanics of the human-AI interaction, specifically AI autonomy and human refinement, as opposed to the external academic setting (e.g., assessment type). While prior work in co-creative writing  suggests attribution is highly nuanced and dependent on the specific details of the contribution \cite{he2025exploring, he2025contributions}, our results clarify which factors define these norms in programming. Consequently, students seem to perceive ownership as a functional property earned through process, contrasting with norms that define authorship solely by the source of generation. For educators and policymakers, this signals a potential misalignment between student intuition and institutional policies relying on binary definitions of ``originality.'' To better define the boundaries of AI use, guidelines might benefit from explicitly addressing the role of refinement, clarifying the threshold where functional modification transforms AI-generated artifact into valid student contribution.

\paragraph{The Gap Between Ethical Beliefs and Disclosure Behavior}
A critical disconnect emerged between the policies students demand and their personal intent to disclose. While students consistently advocated for stricter attribution requirements when AI compromised authorship or grading fairness, these internal ethical judgments did not predict their own likelihood of disclosure. This disconnect suggests that personal adherence to attribution norms is inhibited by external pressures, possibly due to fear of academic penalties or perceived punitive grading. Since our results indicate that students derive a sense of ownership from their refinement efforts, we hypothesize that reframing disclosure as a record of this process, rather than a confession of AI use, could potentially mitigate the fear of liability and encourage transparent disclosure on AI.

\paragraph{\textbf{Implication: Towards Process-Oriented Attribution}}
These findings indicate that existing attribution frameworks, derived from professional practice and prioritizing credit allocation, may be insufficient for educational contexts. Unlike traditional information sources, which are often passively consumed, generative AI often functions as an active workspace that requires verification and integration; consequently, standard citation styles appear ill-suited to this distinct workflow. We propose a shift towards \textit{Process-Oriented Attribution}. Rather than focusing on static acknowledgments, guidelines could prompt students to document the evolution of their solution, explicitly distinguishing between verbatim adoption and meaningful modification. By operationalizing attribution as a cognitive forcing function~\cite{buccinca2021trust}, educators can transform AI disclosure into a pedagogical tool that safeguards student agency and ensures critical engagement with increasingly advanced AI systems.
\section{Limitations and Future Work}

Our study relies on hypothetical vignettes, which allow for factorial manipulation but may not fully capture behavior under the pressure of real grading consequences. Additionally, participants may have reported idealized attribution standards due to social desirability bias. Our sample also reflects Western academic norms; future work should explore how these perceptions vary across different educational cultures. Finally, we focused solely on student perceptions. As attribution policies are enforced by faculty, future research must assess instructor expectations to identify potential misalignments between student norms and institutional requirements.


\begin{acks}
We acknowledge and thank the support of the Natural Sciences and Engineering Research Council of Canada (NSERC), Discovery Grant \#RGPIN-2024-04348.
\end{acks}

\bibliographystyle{ACM-Reference-Format}


\bibliography{reference}
\balance
\appendix
\onecolumn
\section{Appendix}
\subsection{Factorial and Vignette Design}
\label{appendix:factorial}

\begin{table*}[h]
\centering
\small
\renewcommand{\arraystretch}{1.2}
\begin{tabularx}{\textwidth}{l l l >{\RaggedRight\arraybackslash}X}
\toprule
\textbf{Factor} & \textbf{Code} & \textbf{Level Name} & \textbf{Definition \& Criteria} \\
\midrule

\multirow{3}{*}{\textbf{\shortstack[l]{A. Assessment\\Type}}} 
 & A1 & Concept \& Design & Pre-implementation artifacts (Problem framing, architecture, UML) \cite{ali2007design}. Excludes full implementation. \\
 & A2 & Implementation (Solo) & Individual work (scripts, notebooks, SQL) focusing on functional correctness \cite{zastre2019jupyter, manzoor2020auto}. \\
 & A3 & Project / Team & Collaborative multi-sprint work (PRs, CI/CD, coordination) \cite{tafliovich2015evaluation, hundhausen2021evaluating}. \\
\midrule

\multirow{4}{*}{\textbf{\shortstack[l]{B. AI Assistance\\Level}}} 
 & B1 & None / IDE-only & Standard linter/autocomplete only. \textit{No generative AI outputs used.} \\
 & B2 & Q\&A Explanations & Concepts/error interpretation via Chatbot/CodeAid \cite{kazemitabaar2024codeaid}. \textit{No AI text/code is pasted}. \\
 & B3 & Local Generation & Small, localized artifacts ($\leq$30 lines) via tools like Copilot Autocomplete \cite{kazemitabaar2023novices}. \\
 & B4 & Systemic/Agentic & Multi-file orchestration via Agentic IDEs (e.g., Cursor, Windsurf). \\
\midrule

\multirow{3}{*}{\textbf{\shortstack[l]{C. Activity\\Type}}} 
 & C1 & Conceptual / Planning & Task decomposition or design choices without prescribing specific edits. \\
 & C2 & Production & Authoring or substantively modifying the code/text constituting the deliverable. \\
 & C3 & Review / QA & Testing, debugging, refactoring, or style checking. \\
\midrule

\multirow{2}{*}{\textbf{\shortstack[l]{D. Prior\\Knowledge}}} 
 & D1 & From-scratch & Student lacks working solution/knowledge (``help me do this''). \\
 & D2 & Already-passing & Student possesses sufficient knowledge; seeks optimization/extension. \\
\midrule

\multirow{2}{*}{\textbf{\shortstack[l]{E. Human\\Post-AI Effort}}} 
 & E1 & Used As-Is & Minimal Effort: Student accepts AI output with little modification. \\
 & E2 & Significantly Modified & Substantial Effort: Iterative prompting, debugging, or rewriting. \\
\bottomrule
\end{tabularx}
\caption{Factorial Design Definition. Citations indicate relevant work used to exemplify factor levels.}
\label{tab:study_design}
\end{table*}
\vspace{-1em}
Given the five unique factors identified, a full cross product results in 144 unique combinations, 42 of which were removed due to logical incoherence, (e.g., Non-AI (B1) with any Human Post-AI Effort (E1 or E2)), resulting in a total of 102 unique scenarios (vignettes), each with a compelling, easy-to-understand, and generally applicable narrative that encapsulates the variable assigned. The full design spreadsheet can be accessed via \href{https://docs.google.com/spreadsheets/d/1Wk3OlUS9Dhs3FzNXIICcf9JoHJYUBsJkBut71GM5ZMQ/edit?gid=114402785#gid=114402785}{Link to Google Sheets}.

To make the text-based vignette more easily comprehensible, we have also created slide shows accompanying each vignette. The slideshow, which was presented alongside the text-based vignette, can be accessed via \href{https://github.com/hivelabuoft/ai-attribution-in-cs/tree/main/pages}{Link to GitHub Repository}.

\subsection{Survey Design}
For each of the vignettes, participants responded to the following items, grouped by our dependent measures:

\paragraph{Perceived Authorship and Agency.}
\begin{enumerate}
\item \textbf{Work Attribution:} To what extent does the \textbf{work} in this submission reflect human versus AI effort? \\
\textit{[7-point Likert: Entirely Human (1) -- Entirely AI (7)]}
\item \textbf{Intellectual Contribution:} To what extent do the \textbf{new ideas} in this submission reflect human versus AI contribution? \\
\textit{[7-point Likert: Entirely Human (1) -- Entirely AI (7)]}
\item \textbf{Initiative:} Who took more initiative in \textbf{advancing} the work? \\
\textit{[7-point Likert: Entirely Human (1) -- Entirely AI (7)]}
\item \textbf{Liability:} If the final work contained an \textbf{error}, which party would be \textbf{responsible}? \\
\textit{[7-point Likert: Entirely Student (1) -- Entirely AI (7)]}
\end{enumerate}
\paragraph{Pedagogical Validity.}
\begin{enumerate}
\setcounter{enumi}{4}
\item \textbf{Grading Fairness:} How \textbf{fair} would it be to \textbf{grade} this submission as typical for this assessment? \\
\textit{[7-point Likert: Very Unfair (1) -- Very Fair (7)]}
\item \textbf{Student Learning:} How much \textbf{learning} do you believe occurred for the \textbf{student} in the scenario described? \\
\textit{[7-point Likert: None at all (1) -- A great deal (7)]}
\end{enumerate}
\paragraph{Disclosure Preferences.}
\begin{enumerate}
\setcounter{enumi}{6}
\item \textbf{Attribution Intensity:} What level of disclosure regarding AI usage should the student be required to provide in this submission? \\
\textit{[Single select: None, Brief acknowledgment, Citation, Co-authorship]}
\item \textbf{Personal Willingness:} Would you personally disclose AI use if you were this student? (Assuming you will not be penalized or rewarded either way). \\
\textit{[Single-select: Yes / No]}
\end{enumerate}

\subsection{Vignette Design}
\begin{figure}[h]
\centering
\begin{tcolorbox}[
  enhanced,
  width=\columnwidth,
  colback=white,
  colframe=black!40,
  boxrule=0.4pt,
  arc=3pt,
  left=6pt, right=6pt, top=6pt, bottom=6pt,
  fonttitle=\bfseries\small,
  coltitle=white,
  colbacktitle=black!75,
  title=Vignette: \#116
]
 
\small
\hlA{In a second-year programming course, a student was responsible for the file I/O module in a \textbf{team project}.}
\hlD{Their program was \textbf{already functioning correctly},}
\hlC{but the team needed to add \textbf{comprehensive error handling}.}
\hlB{To understand the correct approach, the student \textbf{asked an AI assistant}, ``What is the standard way to handle file-not-found errors in Python?''
The AI \textbf{explained} the} {\ttfamily\hlB{try\dots except FileNotFoundError}} \hlB{block.}
\hlE{Using the explanation as a starting point, the student then \textbf{manually designed and implemented} a more robust solution \textbf{on their own}, creating a multi-part block to also handle} {\ttfamily\hlE{PermissionError}} \hlE{and other potential} {\ttfamily\hlE{IOError}} \hlE{exceptions with \textbf{distinct user messages} for each.}
 
\medskip
\noindent\rule{\linewidth}{0.3pt}
\smallskip
 
\footnotesize\centering
\begin{tabular}{@{}l@{\quad}l@{\quad}l@{}}
\swatch{bgA}\;\textcolor{colA}{\textbf{(A)}}\,Assessment Type &
\swatch{bgB}\;\textcolor{colB}{\textbf{(B)}}\,AI Assistance Level &
\swatch{bgC}\;\textcolor{colC}{\textbf{(C)}}\,Activity Type \\[3pt]
\swatch{bgD}\;\textcolor{colD}{\textbf{(D)}}\,Prior Knowledge &
\swatch{bgE}\;\textcolor{colE}{\textbf{(E)}}\,Post-AI Effort &
\end{tabular}
 
\end{tcolorbox}
 
\caption{Example vignette with factor assignments color-coded.}
\label{fig:vignette-example}
 
\end{figure}

\newpage
\subsection{Study Results}

\begin{table*}[ht]
\centering
\setlength{\tabcolsep}{4pt}
\renewcommand{\arraystretch}{1.2}
\begin{tabular}{llcccccc}
\toprule
Factor & df &
Work Attr. &
Idea Contr. &
Initiative &
Error Resp. &
Grading Fair. &
Student Learn. \\
\midrule
A: Assessment type & 2 &
0.70 (.50) & 0.40 (.67) & 0.02 (.98) & 1.20 (.30) & 0.33 (.72) & 1.20 (.30) \\

B: AI assistance level & 3 &
\cellcolor{Bittersweet!70} 90.91 ($<.001$) &
\cellcolor{Bittersweet!68} 86.30 ($<.001$) &
\cellcolor{Bittersweet!60} 69.85 ($<.001$) &
\cellcolor{Bittersweet!25} 22.85 ($<.001$) &
\cellcolor{Bittersweet!93} 159.02 ($<.001$) &
\cellcolor{Bittersweet!100} 185.67 ($<.001$) \\

C: Activity type & 2 &
0.18 (.83) & 0.32 (.72) & 1.88 (.15) & 0.83 (.44) & 0.87 (.42) & 0.32 (.73) \\

D: Prior knowledge & 1 &
0.00 (.99) & 0.37 (.54) & 0.34 (.56) & 0.05 (.83) & 1.52 (.22) & 1.36 (.24) \\

E: Human post-AI effort & 1 &
\cellcolor{Bittersweet!10} 16.83 ($<.001$) &
\cellcolor{Bittersweet!10} 15.53 ($<.001$) &
\cellcolor{Bittersweet!10} 14.60 ($<.001$) &
1.54 (.22) &
0.98 (.32) &
\cellcolor{Bittersweet!12} 19.35 ($<.001$) \\
\bottomrule
\end{tabular}
\caption{Main-effects factorial ANOVA for student perceptions by experimental factor. Cells show $F$-statistics (with $p$-values in parentheses), background shading is proportional to partial eta squared ($\eta^2_p$), darker color means larger effect size. Residual degrees of freedom were 389 for all models.}
\label{tab:anova_main_effects}
\end{table*}

\begin{figure*}[h]
    \centering
    \includegraphics[width=.8\linewidth]{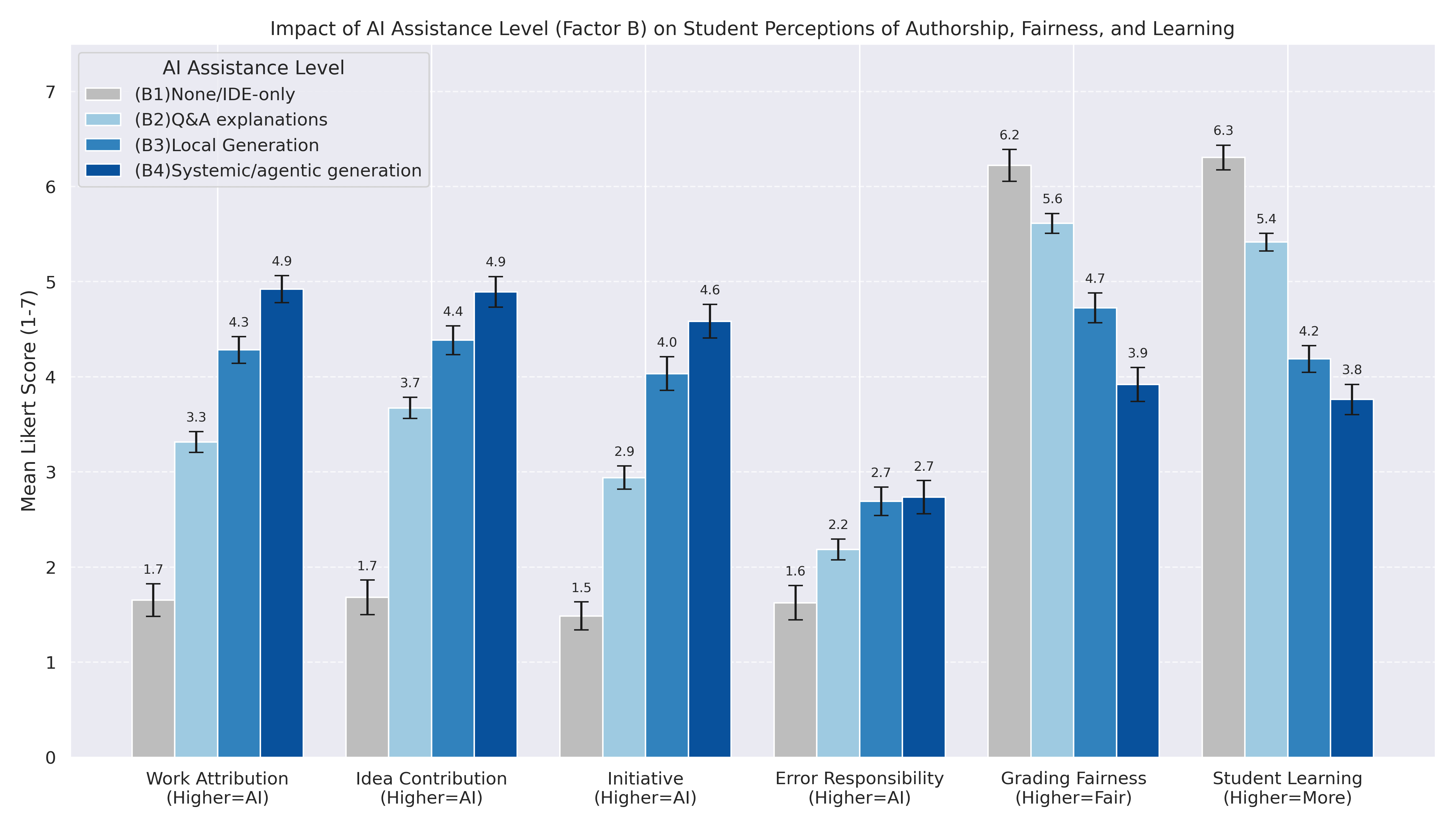}
    \caption{Main effect of AI assistance level (factor B) on mean Likert ratings for authorship, responsibility, grading fairness, and learning, collapsed across other factors; error bars show standard error (SEM).}
    \label{fig:B-ai-assistance-impact-likert}
\end{figure*}
\newpage

\begin{figure*}[h]
    \centering
    \includegraphics[width=.8\linewidth]{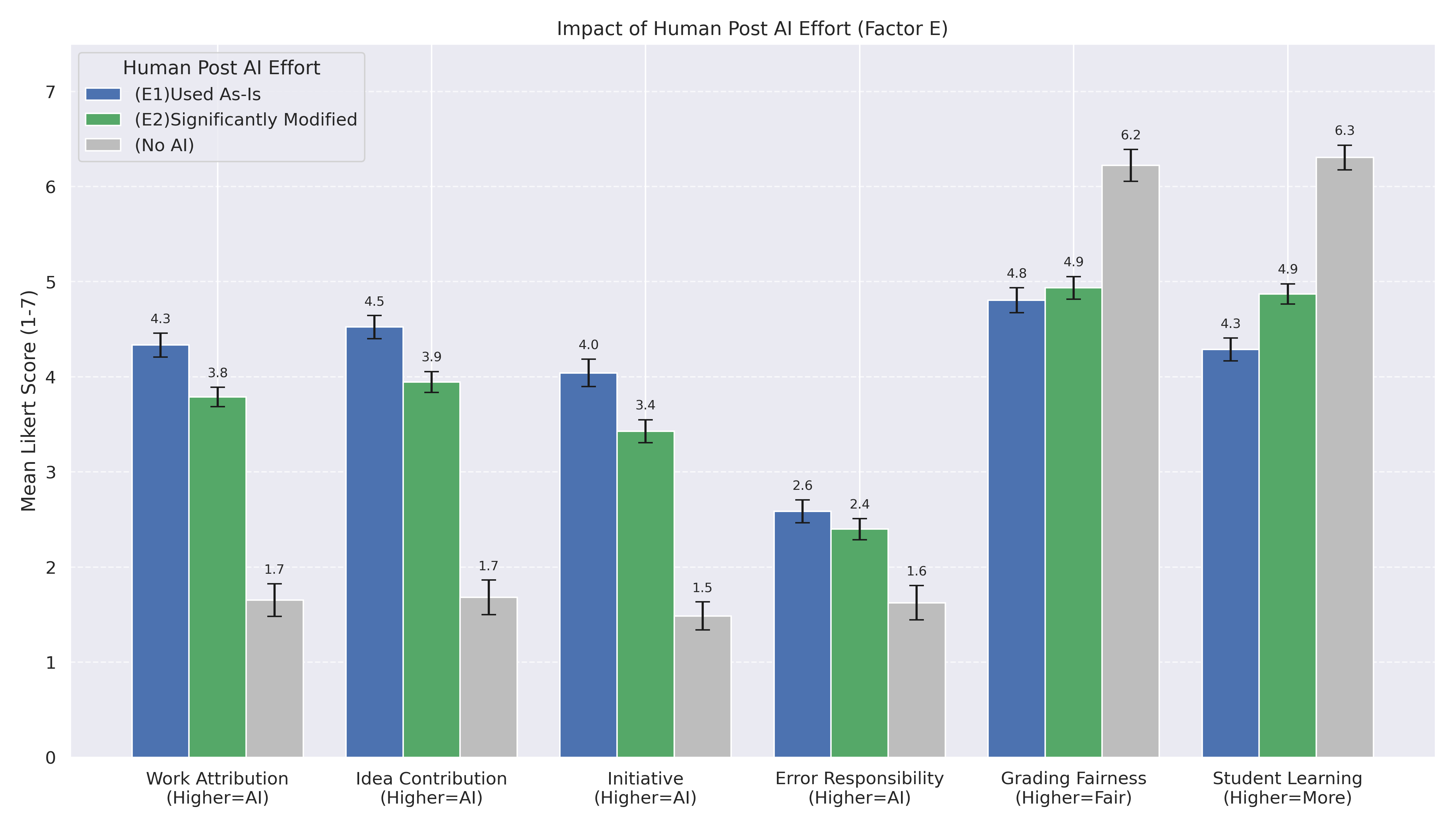}
    \caption{Main effect of human post-AI effort (factor E) on mean Likert ratings for authorship, responsibility, grading fairness, and learning, collapsed across other factors; error bars show standard error (SEM).}
    \label{fig:E-post-ai-human-effort-impact-likert}
\end{figure*}

\begin{table}[ht]
    \centering
    \begin{tabular}{l c c c c}
    \toprule
    \textbf{Perception Metrics} & \textbf{Coef ($\beta$)} & \textbf{Std. Err} & \textbf{z} & \textbf{P$>|z|$} \\
    \midrule
    Work Attribution & 0.276 & 0.105 & 2.619 & \textbf{0.009} \\
    Idea Contribution & 0.246 & 0.097 & 2.550 & \textbf{0.011} \\
    Initiative & 0.157 & 0.084 & 1.883 & 0.060 \\
    Error Responsibility & -0.088 & 0.065 & -1.355 & 0.176 \\
    Grading Fairness & -0.379 & 0.078 & -4.890 & \textbf{< .001} \\
    Student Learning & -0.042 & 0.087 & -0.484 & 0.628 \\
    \bottomrule
    \end{tabular}
    \caption{Ordered Logistic Regression Results for Disclosure Requirement. The model predicts the required level of disclosure (None $\to$ Co-authorship) based on student perceptions (Log-Likelihood = -432.07).}
    \label{tab:disclosure_req_result}
\end{table}
\begin{table}[ht]
    \centering
    \begin{tabular}{l c c c c}
    \toprule
    \textbf{Perception Metrics} & \textbf{Coef ($\beta$)} & \textbf{Std. Err} & \textbf{z} & \textbf{P$>|z|$} \\
    \midrule
    Work Attribution & 0.019 & 0.106 & 0.176 & 0.861 \\
    Idea Contribution & 0.114 & 0.098 & 1.155 & 0.248 \\
    Initiative & 0.012 & 0.080 & 0.143 & 0.886 \\
    Error Responsibility & -0.041 & 0.062 & -0.656 & 0.512 \\
    Grading Fairness & -0.075 & 0.076 & -0.982 & 0.326 \\
    Student Learning & -0.146 & 0.089 & -1.645 & 0.100 \\
    \bottomrule
    \end{tabular}
    \caption{Ordered Logistic Regression Results for Personal Disclosure Preference. The model predicts the students' personal preference on disclosure (No $\to$ Yes) based on student perceptions (Log-Likelihood = -399.47).}
    \label{tab:disclosure_pref_result}
\end{table}
\end{document}